\documentclass[conference]{IEEEtran}
\IEEEoverridecommandlockouts

\usepackage{amsthm,amssymb}
\usepackage{graphicx}
\usepackage{braket}
\usepackage[dvipsnames]{xcolor}
\usepackage{arydshln}
\usepackage[utf8]{inputenc} 
\usepackage[T1]{fontenc}
\usepackage{url}
\usepackage{ifthen}
\usepackage{cite}
\usepackage[cmex10]{amsmath} 
\usepackage{bm}
\usepackage{amsfonts}
\usepackage{xcolor}
\usepackage{bbm}
\usepackage{pgfplots}
\usepackage{colortbl}
\usepackage{pgfplots}
\usepackage{tikz}
\usepackage{tikzpeople}

\newtheorem{theorem}{Theorem}

\newtheorem{remark}{Remark}
\newtheorem{lemma}{Lemma}
\newtheorem{example}{Example}

\newtheorem{claim}{Claim}
\newenvironment{Proof}[1]{\medskip\par\noindent{\bf Proof.\,}\,#1}{{\mbox{\,$\blacksquare$}\par}}

\newcommand{\defeq}{\stackrel{\Delta}{=}}

\pgfplotsset{compat=1.18}
\usetikzlibrary{tikzmark, shapes, fit,backgrounds}
\usetikzlibrary{matrix, decorations.pathreplacing, angles, quotes}
\usetikzlibrary{patterns, calc, positioning}
\usetikzlibrary{arrows, arrows.meta}
\usetikzlibrary{shapes.multipart}
\usetikzlibrary{fit}
\usetikzlibrary{shapes.geometric, positioning}
\usetikzlibrary{decorations.pathmorphing}
\tikzstyle{block} = [rectangle, draw, minimum width=4em, text centered, rounded corners, minimum height=1em]
\tikzstyle{line} = [draw, -latex]
\tikzset{meter/.append style={draw, inner sep=10, rectangle, font=\vphantom{A}, minimum width=30, scale=.7, path picture={\draw[black] ([shift={(.1,.3)}]path picture bounding box.south west) to[bend left=50] ([shift={(-.1,.3)}]path picture bounding box.south east);\draw[black,-{Latex[scale=.5]}] ([shift={(0,.1)}]path picture bounding box.south) -- ([shift={(.3,-.1)}]path picture bounding box.north);}}}
\tikzset{snake it/.style={decorate, decoration=snake}}
\usetikzlibrary{shapes.multipart}
\usetikzlibrary{fit}
\usetikzlibrary{shapes.geometric,positioning}
\usetikzlibrary{calc}
\usetikzlibrary{patterns}
\usetikzlibrary{matrix}
\usetikzlibrary{tikzmark}

\begin{document}

\title{On the Capacity Region of Individual Key Rates in Vector Linear Secure Aggregation}

\author{Lei Hu \quad Sennur Ulukus\\
    \normalsize Department of Electrical and Computer Engineering\\
    \normalsize University of Maryland, College Park, MD 20742 \\
    \normalsize \emph{leihu@umd.edu} \quad \emph{ulukus@umd.edu}}

\maketitle

\begin{abstract}
    We provide new insights into an open problem recently posed by Yuan-Sun [ISIT 2025], concerning the minimum individual key rate required in the vector linear secure aggregation problem. Consider a distributed system with $K$ users, where each user $k\in [K]$ holds a data stream $W_k$ and an individual key $Z_k$. A server aims to compute a linear function $\mathbf{F}[W_1;\ldots;W_K]$ without learning any information about another linear function $\mathbf{G}[W_1;\ldots;W_K]$, where $[W_1;\ldots;W_K]$ denotes the row stack of $W_1,\ldots,W_K$. The open problem is to determine the minimum required length of $Z_k$, denoted as $R_k$, $k\in [K]$. In this paper, we characterize a new achievable region for the rate tuple $(R_1,\ldots,R_K)$. The region is polyhedral, with vertices characterized by a binary rate assignment $(R_1,\ldots,R_K) = (\mathbbm{1}(1 \in \mathcal{I}),\ldots,\mathbbm{1}(K\in \mathcal{I}))$, where $\mathcal{I}\subseteq [K]$ satisfies the \textit{rank-increment condition}: $\mathrm{rank}\left(\bigl[\mathbf{F}_{\mathcal{I}};\mathbf{G}_{\mathcal{I}}\bigr]\right) =\mathrm{rank}\bigl(\mathbf{F}_{\mathcal{I}}\bigr)+N$. Here, $\mathbf{F}_\mathcal{I}$ and $\mathbf{G}_\mathcal{I}$ are the submatrices formed by the columns indexed by $\mathcal{I}$. Our results uncover the novel fact that it is not necessary for every user to hold a key, thereby strictly enlarging the best-known achievable region in the literature. Furthermore, we provide a converse analysis to demonstrate its optimality when minimizing the number of users that hold keys.
\end{abstract}

\section{Introduction}
Motivated by the widespread applications of federated learning in computer science and various fields of engineering \cite{Schlegel2023,so2022lightsecagg,konevcny2016federated,ulukusPIRLC}, the information-theoretical limits on the secure summation problem have been extensively investigated \cite{Wan_Uncoded,zhao2023secure}. Specifically, consider a setting with $K$ distributed users, where user $k \in [K]$ holds a data stream $W_k$ and a key $Z_k$. A server aims to compute the summation of these data streams, $W_1 + \ldots + W_K$, without obtaining any additional information about the messages. To this end, each user $k$ sends message $X_k$ to the server, which is a function of $W_k$ and $Z_k$. The information-theoretic limits of the communication cost and key rate were characterized in \cite{zhao2023secure}: in order to compute one symbol of the summation, each user must transmit one symbol and possess one symbol key, while the total randomness (joint entropy) of all keys should be at least $K-1$.

Recently, a generalization of the secure summation problem called \textit{vector linear secure aggregation} was studied by Yuan and Sun \cite{yuan2025vector}, where the server aims to obtain an arbitrary linear function $\mathbf{F}[W_1;\ldots;W_K]$, where $\mathbf{F} \in \mathbb{F}_q^{M \times K}$ has full row rank $\mathrm{rank}(\mathbf{F}) = M$, and $[W_1;\ldots;W_K]$ denotes the row stack of $W_1,\ldots,W_K$. Meanwhile, the server must not gain any information about another linear function $\mathbf{G}[W_1;\ldots;W_K]$, where $\mathbf{G} \in \mathbb{F}_q^{N \times K}$ with $\mathrm{rank}(\mathbf{G}) = N$. This formulation encompasses some existing problems as special cases, including distributed linear transforms \cite{dutta2016short} and linear network computation \cite{Bai_ISIT2024,Guang_PartI,xu2022network}.

For vector linear secure aggregation, the communication cost and the minimum joint entropy of the keys were fully characterized \cite{yuan2025vector}: for each aggregation, each user must send at least $1$ symbol message, and the minimum joint entropy of the keys is $\mathrm{rank}([\mathbf{F};\mathbf{G}]) - \mathrm{rank}(\mathbf{F})$. However, the optimal individual key rate, $R_k$, i.e., the minimum length of $Z_k, k\in [K]$, is still unknown. Due to the structural complexity of the matrices $\mathbf{F}$ and $\mathbf{G}$, determining these individual rates is substantially more challenging than characterizing the total amount of randomness.

A key difficulty arises from the fact that individual keys can be correlated. The tuple $(R_1,\ldots,R_K)$ forms a region rather than a single operating point. To illustrate this phenomenon, consider in $\mathbb{F}_3$, the server wants to obtain $W_1+W_2+W_3$ without gaining any information about $W_1 + W_3$, i.e., $\mathbf{F}=[1\ 1\ 1]$ and $\mathbf{G} = [1\ 0\ 1]$. The individual key of either user $1$ or $3$ can be zero, but not both simultaneously. For example, an achievable scheme attaining $R_1 = 0$ is given by $X_1 = W_1, X_2 = W_2 + S, X_3 = W_3 - S$, as shown in Example~\ref{example1} in Section~\ref{sect:examples}. Despite such observations, the capacity region of the tuple $(R_1,\ldots,R_K)$ has not yet been explicitly characterized. While Yuan-Sun~\cite{yuan2025vector} proposed an achievable aggregation scheme that characterizes the minimal joint entropy of the keys, their result does not yield a transparent correspondence to individual key rates and appears to suggest the condition $R_k \geq 1$ for all $k\in [K]$ is necessary. 

In this work, we show that this condition is in fact not necessary: not all users are required to hold individual keys. We characterize a new achievable region in the form of a polyhedron whose vertices are $(R_1,\ldots,R_K) = (\mathbbm{1}(1 \in \mathcal{I}),\ldots,\mathbbm{1}(K\in \mathcal{I}))$, where $\mathcal{I} \subseteq [K]$ is any index set satisfying the \textit{rank-increment condition}:
\begin{align}
    \mathrm{rank}\left(\bigl[\mathbf{F}_{\mathcal{I}};\mathbf{G}_{\mathcal{I}}\bigr]\right) =\mathrm{rank}\bigl(\mathbf{F}_{\mathcal{I}}\bigr)+N,
\end{align}
where $\mathbf{F}_{\mathcal{I}}$ and $\mathbf{G}_{\mathcal{I}}$ are matrices formed by the columns indexed by $\mathcal{I}$.
We prove that each such vertex is achievable via a linear coding scheme, where an encoding matrix is adopted to map the source keys to the individual keys. Furthermore, we establish a converse bound showing that these vertices are the corner points of the capacity region, and are therefore optimal in terms of minimizing the number of users that must store individual keys. This result reveals a previously unexplored asymmetry in the role of users in vector linear secure aggregation, and strictly enlarges the best-known achievable region.

\textit{Notation:}
For positive integer $M$, $[M]$ denotes the set $\{1, \ldots, M\}$. For integers $a$ and $b$, $[a:b]$ denotes $\{a,\ldots,b \}$. $A_{[M]}$ is the compact notation of $\{A_{1}, A_2, \ldots, A_{M}\}$. $\mathbb{F}_{q}$ denotes a finite field with order $q$, where $q = p^r$ with $p$ a prime number and $r$ a positive integer. The symbol $\defeq$ means ``defined as''. $\mathbb{F}_{q}^{a \times b}$ is the set of $a \times b$ matrices with elements in $\mathbb{F}_{q}$. $\mathbf{I}_{a}$ is an $a \times a$ identity matrix, and $\mathbf{0}_{a \times b}$ is the $a \times b$ all-zero matrix. $\mathbbm{1}(\cdot)$ denotes the indicator function. $\mathsf{conv}(\cdot)$ denotes convex hull. For a matrix $\mathbf{A}$, $\mathrm{rowspan}(\mathbf{A})$ and $\mathrm{colspan}(\mathbf{A})$
denote the vector subspace spanned by the rows and columns of $\mathbf{A}$, respectively. For a set $\mathcal{I}$, $|\mathcal{I}|$ denotes its cardinality. The notation $\mathcal{I} \setminus {i}$ refers to the set obtained by removing the element $i$ from $\mathcal{I}$, and $\mathcal{I} \setminus \mathcal{J}$ denotes the set obtained by removing all elements of the set $\mathcal{J}$ from $\mathcal{I}$.

\section{Problem Statement}
\subsection{System Model}
We consider a distributed system consisting of $K$ users and a server. Each user $k \in [K]$ holds a data stream with length $L$, $W_k \in \mathbb{F}_q^{1 \times L}$, and an individual key vector $Z_k \in \mathbb{F}_q^{1 \times L_k}$ with length $L_k$. Each data stream $W_k$ is uniformly distributed over $\mathbb{F}_q$, and the data streams $W_{[K]}$ are independent of the keys $Z_{[K]}$, i.e.,
\begin{align}
    H(W_k) & = L, \ \forall k \in [K], \\
    H(W_{[K]}, Z_{[K]}) & = \sum_{k \in [K]} H(W_k) + H(Z_{[K]}).
\end{align}
Moreover, all individual keys $Z_{[K]}$ are generated from a common source key $Z_{\Sigma} \in \mathbb{F}_q^{1 \times L_{\Sigma} }$ with length $L_{\Sigma}$, i.e.,
\begin{align}
    H(Z_{[K]}|Z_{\Sigma}) = 0.
\end{align}

The server's goal is to compute a linear function $F$ of the data streams, while obtaining no information about another protected linear function $G$. To be specific, let $\mathbf{W} \defeq [W_1;\ldots ; W_K] \in \mathbb{F}_q^{K \times L}$ be the row stack of the data streams. The server aims to compute
\begin{align}
    F \defeq \mathbf{F} \mathbf{W} \in \mathbb{F}_q^{M \times L},
\end{align}
where $\mathrm{rank}(\mathbf{F}) = M$. At the same time, the server must not get any information about
\begin{align}
    G \defeq \mathbf{G} \mathbf{W} \in \mathbb{F}_q^{N \times L},
\end{align}
where $\mathrm{rank}(\mathbf{G}) = N$.
Without loss of generality, we make the following assumptions:
\begin{enumerate}
    \item $\mathrm{rank}([\mathbf{F};\mathbf{G}]) = M + N$, ensuring that the desired computation $F$ is linearly independent of the protected computation $G$; otherwise, it is impossible to fully protect $G$.
    \item Each column of $\mathbf{F}$ contains at least one nonzero entry; if a column is all zeros, the corresponding data stream is unused and can be removed from the model.
\end{enumerate}

\subsection{Communication Process}
The communication process for vector secure aggregation proceeds as follows. First, each user $k \in [K]$ sends a message $X_k$ with length $L_X$ to the server, where $X_k$ is a function of its local data stream $W_k$ and individual key $Z_k$, i.e.,
\begin{align}
    H(X_k|W_k,Z_k) = 0, \ \forall k \in [K].
\end{align}

Upon receiving all messages $X_{[K]}$, the server must be able to recover the desired computation, i.e.,
\begin{align}
   \text{[Correctness]} \quad   H(F|X_{[K]}) = 0,
\end{align}
while no information about $G$ is revealed to the server:
\begin{align}
   \text{[Security]} \quad  I(G;X_{[K]}|F) = 0.
\end{align}

\subsection{Communication and Key Rates}
We now define three metrics to evaluate the communication cost and key distribution in this model.
\begin{itemize}
    \item \textbf{Communication rate:} defined as the number of symbols sent by each user per aggregation,
    \begin{align}
        R_X \defeq \frac{L_X}{L}. \label{eq:R_X}
    \end{align}

    \item \textbf{Total key rate:} defined as the joint entropy (total randomness) of the keys, 
    \begin{align}
        R_{Z_{\Sigma}} \defeq \frac{L_{Z_{\Sigma}}}{L}.\label{eq:R_Z}
    \end{align}

    \item \textbf{Individual key rate tuple:} specifies the length of individual keys per aggregation,
    \begin{align}
        (R_1, R_2, \ldots, R_K) \defeq \left(\frac
        {L_1}{L}, \frac
        {L_2}{L}, \ldots, \frac
        {L_K}{L} \right).
    \end{align}
    An individual key rate tuple $(R_1,\ldots,R_K)$ is called \textit{achievable} if there exists a communication scheme and block length $L$ such that the correctness and security constraints are satisfied using the individual key tuple $(L R_1,\ldots, LR_K)$. Furthermore, a region $\mathfrak{R}_{\mathsf {achi}}$ is called achievable if every tuple in this region is achievable. Finally, the \textit{capacity region} is defined as the closure of the set of all achievable individual key rate tuples.
\end{itemize}

Yuan-Sun \cite{yuan2025vector} showed that the optimal (minimum) communication rate $R_X^*$ is $1$ and the optimal total key rate $R_{Z_\Sigma}^*$ is $N$ (assuming  $\mathrm{rank}([\mathbf{F};\mathbf{G}]) = M + N$). However, their work does not provide a clear characterization of an achievable region for the individual key rate tuples. 
In this paper, we introduce a new achievable region $\mathfrak{R}_{\mathsf{achi}}$ and demonstrate its optimality in certain cases.

\section{Main Results}
\begin{theorem}\label{thm-1}
    Let $\mathbf{F} \in \mathbb{F}_q^{M \times K}$ and $ \mathbf{G} \in \mathbb{F}_q^{N \times K}$ be any two matrices satisfying $\mathrm{rank}\left( \bigl[\mathbf{F};\mathbf{G}\bigr] \right) = M + N$.
    The following individual key rate region is achievable,
    \begin{align}
        \mathfrak{R}_{\mathsf{achi}} = \mathsf{conv} \left( 
        \begin{aligned} 
            \bigcup_{\substack{\mathcal{I} \subseteq [K], \\ \ \mathrm{rank}([\mathbf{F}_\mathcal{I}; \mathbf{G}_\mathcal{I}] ) = \mathrm{rank}(\mathbf{F}_\mathcal{I}) + N}} \mathfrak{R}_{\mathcal{I}} 
        \end{aligned}\right),\label{eq:R_I}
    \end{align}
    where 
    \begin{align}
        \mathfrak{R}_{\mathcal{I}} \defeq \left\{ (R_1,\ldots,R_K) \in \mathbb{R}_{+}^{K}  \mid R_i \geq 1, \ \forall i \in \mathcal{I} \right\}.
    \end{align}
\end{theorem}

The proof of Theorem~\ref{thm-1} is given via a general achievability scheme in Section \ref{Sec:achi}. 

\begin{remark}\label{remark-1}
    The region \(\mathfrak{R}_{\mathsf{achi}}\) is a convex polyhedron whose vertices are the vectors $(R_1,\ldots,R_K) = (\mathbbm{1}(k\in \mathcal{I}),\ldots,\mathbbm{1}(K \in \mathcal{I}))$, where the index set $\mathcal{I}$ satisfies the rank-increment condition, $\mathrm{rank}([\mathbf{F}_{\mathcal{I}}; \mathbf{G}_{\mathcal{I}}] ) = \mathrm{rank}(\mathbf{F}_{\mathcal{I}}) + N$. 
    By explicitly constructing a scheme that achieves each such vertex and applying time‐sharing, the entire region $\mathfrak{R}_{\mathsf{achi}}$ is achievable. 
\end{remark}

Since the rank-increment condition plays a crucial role in our results, we next provide further analysis of this condition.

\begin{remark}
    The collection of index sets satisfying the rank-increment condition is upward closed: if $\mathcal{I}$ satisfies the condition and $\mathcal{I} \subseteq \mathcal{J}$, then $\mathcal{J}$ also satisfies the condition and $\mathfrak{R}_{\mathcal{J}} \subseteq \mathfrak{R}_{\mathcal{I}}$. Therefore, the
    inclusion-wise minimal index sets $\mathcal{I}^* \subseteq [K]$ determine the vertices of the region. Here, ``inclusion-wise minimal'' means that removing any element $i \in \mathcal{I}^*$ breaks the condition, i.e., 
    \begin{align}
        \mathrm{rank}([\mathbf{F}_{\mathcal{I}^*\setminus \{i\}};\mathbf{G}_{\mathcal{I}^*\setminus \{i\}}]) < \mathrm{rank}(\mathbf{F}_{\mathcal{I}^*\setminus \{i\}}) + N.
    \end{align}
\end{remark}

\begin{remark}
    The rank-increment condition can equivalently be stated as
    \begin{align}
        \!\!\! \mathrm{rank}(\mathbf{G}_\mathcal{I}) = N, \  \mathrm{rowspan}(\mathbf{F}_{\mathcal{I}}) \cap \mathrm{rowspan}(\mathbf{G}_{\mathcal{I}}) = \{0\},\label{eq:equiv_cond} 
    \end{align}
    which follows from the fact that
    \begin{align}
        \mathrm{rank}([\mathbf{F}_\mathcal{I}; \mathbf{G}_\mathcal{I}] ) \leq \mathrm{rank}(\mathbf{F}_\mathcal{I} ) + \mathrm{rank}(\mathbf{G}_\mathcal{I}) \leq \mathrm{rank}(\mathbf{F}_\mathcal{I}) + N.
    \end{align}
    An operational interpretation of (\ref{eq:equiv_cond}) is as follows. When individual keys are assigned exclusively to users in a set $\mathcal{I}$ (i.e., only data streams in $\mathcal{I}$ are protected), the first condition in (\ref{eq:equiv_cond}) ensures that the protected subspace, $\mathrm{rowspan}(\mathbf{G}_\mathcal{I})$, spans an $N$-dimensional subspace; the second condition guarantees that the information revealed in $\mathrm{rowspan}(\mathbf{F}_{\mathcal{I}})$ is independent of the protected information in $\mathrm{rowspan}(\mathbf{G}_\mathcal{I})$.
\end{remark}

We now analyze the size of index sets $\mathcal{I}$ satisfying the rank-increment condition. First, observe that
\begin{align}
    |\mathcal{I}| \geq \mathrm{rank}([\mathbf{F}_\mathcal{I}; \mathbf{G}_\mathcal{I}]) = N + \mathrm{rank}(\mathbf{F}_\mathcal{I}). \label{eq:size_I}
\end{align}
Moreover, an important observation is that equality in (\ref{eq:size_I}) holds for any inclusion-wise minimal subset $\mathcal{I}^*$; that is, the matrix $[\mathbf{F}_{\mathcal{I}^*}; \mathbf{G}_{\mathcal{I}^*}]$ is full column rank. This is formalized as the following Theorem.

\begin{theorem}\label{thm-I*}
    Any inclusion-wise minimal subset $\mathcal{I}^*$ satisfying the rank-increment condition also satisfies
    \begin{align}
        |\mathcal{I}^*| = N + \mathrm{rank}(\mathbf{F}_{\mathcal{I}^*}).\label{eq:size_equal}
    \end{align}
\end{theorem}

The proof of Theorem~\ref{thm-I*} is presented in Appendix~\ref{proof:thm-I*}.

We next establish a converse result characterizing the vertices of $\mathfrak{R}_{\mathsf{achi}}$.

\begin{theorem}\label{thm:2}
    For an index set $\mathcal{I} \subseteq [K]$, if individual keys are assigned only to users in $\mathcal{I}$, then $\mathcal{I}$ must satisfy the rank-increment condition. 
    Moreover, if $\mathcal{I}$ is also inclusion-wise minimal, then
    \begin{align}
        R_i \geq 1, \quad \forall i \in \mathcal{I}.
    \end{align}
\end{theorem}

The proof of Theorem~\ref{thm:2} is given in Appendix~\ref{proof:thm-2}.

\begin{remark}
    Theorem \ref{thm:2} shows that the rank-increment condition is necessary for the set of users holding the keys.
    Furthermore, when keys are only assigned to users in an inclusion-wise minimal set $\mathcal{I}^*$, each such user must hold one symbol key per aggregation. As a result, the vertices of $\mathfrak{R}_{\mathsf{achi}}$, given by $(R_1,\ldots,R_K) = ( \mathbbm{1}(1 \in \mathcal{I}^*), \ldots,  \mathbbm{1}(K \in \mathcal{I}^*))$, are also corner points of the capacity region.
\end{remark}

\begin{remark}
    As a special case, consider the secure summation problem studied in \cite{zhao2023secure}, where $\mathbf{F}=[1,1,\ldots,1]$ and $\mathbf{G} = \mathbf{I}_K$. In this setting, any row of $\mathbf{G}$ can be removed, as it can be expressed as a linear combination of the remaining rows of $\mathbf{G}$ together with the rows of $\mathbf{F}$.
    By Theorem~\ref{thm-1}, the only index set that satisfies the rank-increment condition is $[K]$, yielding the achievable region
    \begin{align}
        \mathfrak{R}_{\mathsf{achi}} = \mathfrak{R}_{[K]} = \left\{ (R_1,\ldots,R_K) \mid R_k \geq 1, \ \forall k \in [K] \right\}.
    \end{align}
    This coincides exactly with the capacity region characterized in \cite{zhao2023secure}. Hence, the region derived in this work reduces to the known capacity region for the secure summation problem.
\end{remark}
        
\section{Illustrative Examples}\label{sect:examples}
We now present examples to show the description of the achievable region and the design of the corresponding achievability scheme.

\begin{example}\label{example1}
    Consider $K=3$ users with data streams $W_1,W_2, W_3 \in \mathbb{F}_3$. The server wants to compute $W_1 + W_2 + W_3$ while learning no information about $W_1 + W_3$, i.e., the corresponding computation matrices are
    \begin{align}
        \mathbf{F} = 
        \begin{bmatrix}
            1 & 1 & 1
        \end{bmatrix},\quad 
        \mathbf{G} = 
        \begin{bmatrix}
            1 & 0 & 1
        \end{bmatrix}.
    \end{align}
    To apply Theorem 1, the index sets satisfying the rank-increment condition are
    \begin{align}
        \mathcal{I}_1 = \{1,2\}, \quad
        \mathcal{I}_2 = \{2,3\}, \quad 
        \mathcal{I}_3 = \{1,2,3\}.
    \end{align}
    For instance, we can verify the condition for $\mathcal{I}_1$ as
    \begin{align}
        \mathrm{rank}(\mathbf{F}_{\{1,2\}}; \mathbf{G}_{\{1,2\}}) = 2 = \mathrm{rank}\left( \mathbf{F}_{\{1,2\}} \right) + 1.
    \end{align}
    Then, we can construct a linear scheme achieving the tuple $(R_1,R_2,R_3) = (\mathbbm{1}(1 \in \mathcal{I}_1),\mathbbm{1}(2 \in \mathcal{I}_1), \mathbbm{1}(3 \in \mathcal{I}_1) ) = (1,1,0)$. Let $S \in \mathbb{F}_3$ be a uniformly distributed source key independent of $W_{[3]}$. The users transmit
    \begin{align}
        X_1 = W_1 + S, \ X_2 = W_2 - S, \ X_3 = W_3.
    \end{align}
    The correctness of this scheme is verified as
    \begin{align}
        W_1 + W_2 + W_3 = X_1 + X_2 + X_3,
    \end{align}
    and the security constraint is satisfied by
    \begin{align}
        &I(W_1 + W_3; X_{[3]} | W_1 + W_2 + W_3) \nonumber \\
        & =  H(X_{[3]}, W_1 + W_2 + W_3) - H(W_1 + W_2 + W_3) \notag \\
        & \quad - H(X_{[3]}|W_1 + W_2 + W_3, W_1 + W_3) \\
        & = 3 - 1 - H(X_{[3]}, X_1 + X_3| W_1 + W_2 + W_3, W_1+W_3) \label{eq:ineq_X3} \\
        & = 2 - H(X_{[3]}, S| W_1 + W_2 + W_3, W_1+W_3) \\
        & = 2 - H(W_{[3]}, S | W_1 + W_2 + W_3, W_1+W_3) \\
        & = 2 - H(S) - H(W_{[3]} | W_1 + W_2 + W_3, W_1+W_3) \label{eq:S}\\
        & = 2-1-1=0,
    \end{align}
    where (\ref{eq:ineq_X3}) is because $H(X_{[3]},W_1+W_2+W_3) = H(X_{[3]} )+ H(W_1+W_2+W_3 | X_{[3]}) = H(X_{[3]})=3$, and (\ref{eq:S}) is due to the independence between $S$ and $W_{[3]}$. 

    Following the same procedures, the individual key rate tuples  $(0,1,1)$ and $(1,1,1)$ are also achievable.
    Therefore, the following regions containing more randomness are achievable:
    \begin{align}
         \!\!\!\mathfrak{R}_{\mathcal{I}_1} & = \{ (R_1, R_2,R_3) \in \mathbb{R}_+^{3} \mid R_1 \geq 1, R_2 \geq 1\},\\
         \!\!\!\mathfrak{R}_{\mathcal{I}_2} & = \{ (R_1, R_2,R_3) \in \mathbb{R}_+^{3} \mid R_2 \geq 1, R_3 \geq 1\},\\
         \!\!\!\mathfrak{R}_{\mathcal{I}_3} & = \{ (R_1, R_2,R_3) \in \mathbb{R}_+^{3} \mid R_1 \geq 1, R_2 \geq 1, R_3 \geq 1\}. \!
    \end{align}
    Since $\mathfrak{R}_{\mathcal{I}_3} \subseteq \mathfrak{R}_{\mathcal{I}_1}$ and  $\mathfrak{R}_{\mathcal{I}_3} \subseteq \mathfrak{R}_{\mathcal{I}_2}$, the achievable region is given by the convex hull
    \begin{align}
        \mathfrak{R}_{\mathsf{achi}} & = \mathsf{conv}\left( \mathfrak{R}_{\mathcal{I}_1} \cup \mathfrak{R}_{\mathcal{I}_2} \right) \\
        & = \left\{(R_1,R_2,R_3) \in \mathbb{R}_+^3 \middle|
            \begin{array}{l}
            R_2 \geq 1, \\
            R_1 + R_3 \geq 1
            \end{array} \right\}.\label{eq:achi_ex1}
    \end{align}
    The region $\mathfrak{R}_{\mathsf{achi}}$ is illustrated in Fig. \ref{fig:ex1}. 

    \begin{figure}[h]
        \center	
        \begin{tikzpicture}[x={(170+25+10:1cm)},y={(-15-25+10:1cm)},z={(90:1cm)},scale=1.4,>=stealth,transform shape]
    
        \coordinate (A1) at (0,1,1);
        \coordinate (A2) at (1,1,0);
    
        \draw [line join=bevel, draw=black!70, line width=0.3pt,
           left color=blue!50, right color=blue!15,
           shading=axis, shading angle=240]
        (A1) -- (A2) -- (2,1,0) -- (2,1,2) -- (0,1,2) -- cycle;
    
        \draw [line join=bevel, draw=black!70, line width=0.3pt, left color=blue!50, right color=blue!15, shading=axis, shading angle=120] (A1) -- (0,3,1) -- (0,3,2)-- (0,1,2) -- cycle;
    
        \draw [line join=bevel, draw=black!70, line width=0.3pt, left color=blue!35, right color=blue!40, shading=axis, shading angle=225] (A1) -- (0,3,1) --  (1,3,0) -- (1,1,0) -- cycle;
    
        \draw [line join=bevel, draw=black!70, line width=0.3pt, left color=blue!50, right color=blue!15, shading=axis, shading angle=340] (A2) -- (1,3,0) --  (2,3,0)  -- (2,1,0)-- cycle;
    
        \draw [->, semithick, black] (0,0,0)--(2,0,0);
        \draw [->, semithick, black] (0,0,0)--(0,3,0);
        \draw [->, semithick, black] (0,0,0)--(0,0,2);
        \draw [dashed, thin, black] (1,0,0)--(1,1,0);
        \draw [dashed, thin, black] (0,1,0)--(1,1,0);
        \draw [dashed, thin, black] (0,1,0)--(0,1,1);
        \draw [dashed, thin, black] (0,0,1)--(0,1,1);
    
        \node at (2,0,0) [above, scale=0.5]{\large $R_1$};
        \node at (0,3,0) [above, scale=0.5]{\large $R_2$};
        \node at (0,0,2) [left, scale=0.5]{\large $R_3$};
    
        \node at (A1) [above right = 0.2 and 0.1, scale=0.5]{\large $(0,1,1)$};
        \node at (A1) [black!70] { $\cdot$};
    
        \node at (A2) [above left = 0.3 and 0.5, scale=0.5]{\large $(1,1,0)$};
        \node at (A2) [black!70] { $\cdot$};
    
        \end{tikzpicture}
        \caption{The region $\mathfrak{R}_{\mathsf{achi}}$ for Example~\ref{example1}.}\label{fig:ex1}
    \end{figure}
    
    \begin{remark}
        The intuition behind the region can be explained as follows. To satisfy the security requirement, $W_1 + W_3$ must be protected from the server. Meanwhile, the correctness requirement allows the server to recover $W_1 + W_2 + W_3$. As a consequence, $W_2$ should be masked by the individual key at user $2$, which implies $R_2 \geq 1$.
        
        On the other hand, in order for the server to correctly decode the desired sum, the noise injected into $X_2$ must be canceled at the server. This cancellation can only be achieved by the noises contributed by users $1$ and $3$. Therefore, the total amount of noise provided by users $1$ and $3$ must be at least one $1$, which yields the constraint $R_1 + R_3 \geq 1$.
    \end{remark}

    \begin{remark}
        Recall from Remark \ref{remark-1} that the vertices of $\mathfrak{R}_{\mathsf{achi}}$ are induced by inclusion-wise minimal index sets. In this example, these sets are $\mathcal{I}_1$ and $\mathcal{I}_2$. The set  $\mathcal{I}_3$ is not inclusion-wise minimal, since removing either element $1$ or $3$  does not violate the rank-increment condition. Hence, the rate tuples $(1,1,0)$ and $(0,1,1)$ corresponding to  $\mathcal{I}_1$ and $\mathcal{I}_2$, respectively, form the vertices of the region, as shown in Fig.~\ref{fig:ex1}. This example illustrates that when characterizing the achievable region, it suffices to focus on inclusion-wise minimal index sets, rather than all index sets satisfying the rank-increment condition.
    \end{remark}
\end{example}

\begin{example}\label{example2}
    (Example 2 in Yuan-Sun \cite{yuan2025vector}) In $\mathbb{F}_7$, let
    \begin{align}
        \mathbf{F} & = \begin{bmatrix}
            1 & 0 & 5 & 5 & 3 & 5\\
            0 & 1 & 5 & 6 & 0 & 3
        \end{bmatrix}, \\
        \mathbf{G} & = \begin{bmatrix}
            3 & 0 & 1 & 4 & 2 & 4\\
            2 & 2 & 1 & 3 & 5 & 3\\
            1 & 1 & 3 & 4 & 3 & 1
        \end{bmatrix}.
    \end{align}
    Note that the last row of $\mathbf{G}$ is the sum of the rows of $\mathbf{F}$. Hence, this row cannot be protected and may be removed without loss of generality. Moreover, by row operations using the rows of $\mathbf{F}$, the protected computation is equivalent to
    \begin{align}
        \mathbf{G}' = \begin{bmatrix}
            0 & 0 & 0 & 1 & 0 & 1\\
            0 & 0 & 1 & 0 & 3 & 3
        \end{bmatrix}.
    \end{align}
    The inclusion-wise minimal index sets satisfying the rank-increment condition are
    \begin{align}
        \mathcal{I}_1 &= \{1,2,3,4\},\ \mathcal{I}_2 = \{1,2,3,6\},\    \mathcal{I}_3 =\{1,2,4,5\},\notag \\
        \mathcal{I}_4 &= \{1,2,4,6\},\    \mathcal{I}_5 = \{1,2,5,6\}, \  \mathcal{I}_6 = \{1,3,4,5\},  \notag \\
        \mathcal{I}_7 &=  \{1,3,4,6\}, \   \mathcal{I}_8 = \{1,3,5,6\}, \    \mathcal{I}_9 = \{1,4,5,6\}, \notag \\
        \mathcal{I}_{10} & = \{2,3,4,5\}, \  \mathcal{I}_{11}= \{2,3,4,6\}, \  \mathcal{I}_{12}= \{2,3,5,6\},  \notag \\
        \mathcal{I}_{13} & =\{2,4,5,6\}, \  \mathcal{I}_{14}= \{3,4,5,6\}.
    \end{align}
    For the index set $\mathcal{I}_1$, we design a scheme as follows.
    There exists a matrix $\mathbf{P} \in \mathbb{F}_7^{6 \times 2}$ given by
    \begin{align}
        \mathbf{P} = \begin{bmatrix}
            2 & 2 & 1 & 0 & 0 & 0 \\
            2 & 1 & 0 & 1 & 0 & 0
        \end{bmatrix}^T,
    \end{align}
    such that 
    \begin{align}
        \mathbf{F}\mathbf{P} = \mathbf{0}_{2 \times 2}, \quad \mathrm{rank}(\mathbf{G}' \mathbf{P}) = 2.\label{eq:FP0}
    \end{align}
    Then, a linear coding scheme is specified by
    \begin{align}
        \left[\begin{matrix}
        X_1 \\
        \vdots \\
        X_6
    \end{matrix}\right]
        = 
        \left[\begin{matrix}
        W_1 \\
        \vdots \\
        W_6
    \end{matrix}\right]
        + 
    \mathbf{P}
     \left[\begin{matrix}
        S_1 \\
        S_2
    \end{matrix}\right]
    = 
    \left[\begin{matrix}
        W_1 + 2S_1 + 2 S_2 \\
        W_2 + 2S_1 +  S_2 \\
        W_3 + S_1 \\
        W_4 + S_2 \\
        W_5 \\
        W_6
    \end{matrix}\right],\label{eq:scheme-2}
    \end{align}
    where $S_1$ and $S_2$ are independent source keys with uniform distribution in $\mathbb{F}_7$. Define $\mathbf{x} \defeq [X_1;\cdots;X_6]$ and $\mathbf{w} \defeq [W_1;\cdots;W_6]$.
    The correctness of the scheme is verified by
    \begin{align}
        \mathbf{F}\mathbf{x} = \mathbf{F}\mathbf{w} + \underbrace{\mathbf{F}\mathbf{P}}_{= \mathbf{0}}
        \left[
        \begin{matrix}
            S_1 \\
            S_2
        \end{matrix}
        \right]
        =\mathbf{F}\mathbf{w}.
    \end{align}
    Also, the security constraint is satisfied by
    \begin{align}
         &I(\mathbf{G}\mathbf{w}; X_{[6]} | \mathbf{F}\mathbf{w} ) = I(\mathbf{G}'\mathbf{w}; X_{[6]} | \mathbf{F}\mathbf{w} ) \\
        & = H(X_{[6]}, \mathbf{F}\mathbf{w}) - H(\mathbf{F}\mathbf{w}) - H(X_{[6]}, \mathbf{G}' \mathbf{x} | \mathbf{F}\mathbf{w}, \mathbf{G}' \mathbf{w}) \\
        & = 6 - 2 - H(X_{[6]}, \mathbf{G}' \mathbf{P} [S_1;S_2] | \mathbf{F}\mathbf{w}, \mathbf{G}' \mathbf{w}) \\
        & = 4 - H(W_{[6]}, S_1, S_2 | \mathbf{F}\mathbf{w}, \mathbf{G}' \mathbf{w}) \label{eq:X_W} \\
        & = 4 - 2 - H(W_{[6]}| \mathbf{F}\mathbf{w}, \mathbf{G}'\mathbf{w}) = 0,
    \end{align}
    where (\ref{eq:X_W}) is because $\mathrm{rank}(\mathbf{G}' \mathbf{P}) = 2$ in (\ref{eq:FP0}).
    
    For each index set $\mathcal{I}_j$, $j\in [14]$, an achievable region $\mathfrak{R}_{\mathcal{I}_j}$ is defined according to the rate assignment in (\ref{eq:R_I}).
    By time-sharing among these schemes, the achievable region is characterized by
    \begin{align}
        \mathfrak{R}_{\mathsf{achi}} & = \mathsf{conv}\left( \mathfrak{R}_{\mathcal{I}_1} \cup \mathfrak{R}_{\mathcal{I}_2}\cup
        \cdots \cup \mathfrak{R}_{\mathcal{I}_{14}} \right).
    \end{align}

    \begin{remark}
        For this example, the scheme in Yuan-Sun \cite{yuan2025vector} requires $R_k=1$ for all $k\in [6]$, i.e., each user holds a key.
        In contrast, the scheme in~\eqref{eq:scheme-2} achieves $(R_1,\ldots,R_6)=(1,1,1,1,0,0)$, thus not all users need to hold individual keys. Hence, the achievable region $\mathfrak{R}_{\mathsf{achi}}$ obtained here is strictly larger.
    \end{remark}
\end{example}

\section{Achievability Proof}\label{Sec:achi}
\subsection{A Framework for Vector Linear Secure Aggregation }\label{Sec:framework}
We present a general framework for vector secure linear aggregation. Our scheme only requires a single symbol per user, i.e., the data stream length is $L=1$. The messages $X_{[K]}$ sent by the users are constructed as
\begin{align}
    \underbrace{\left[\begin{matrix}
        X_1 \\
        X_2 \\
        \vdots \\
        X_K
    \end{matrix}\right]}_{\defeq \  \mathbf{x}}
    =
    \underbrace{\left[\begin{matrix}
        W_1 \\
        W_2 \\
        \vdots \\
        W_K
    \end{matrix}\right]}_{\defeq \ \mathbf{w}}
    + \underbrace{\left[\begin{matrix}
        Z_1 \\ 
        Z_2 \\
        \vdots \\
        Z_K
    \end{matrix}\right]}_{\defeq \ \mathbf{z}}
    = 
    \mathbf{w}
    + 
    \mathbf{P}
    \underbrace{\left[\begin{matrix}
        S_1 \\ 
        S_2 \\
        \vdots \\
        S_N
    \end{matrix}\right]}_{\defeq \ \mathbf{s}},\label{eq:general_scheme}
\end{align}
where $Z_{[K]}$ are the individual keys, $S_{[N]}$ consists of $N$ independent and uniformly distributed source keys in $\mathbb{F}_q$, and $\mathbf{P} \defeq [\mathbf{p}_1^T;\ldots;\mathbf{p}_K^T] \in \mathbb{F}_q^{K \times N}$ is an encoding matrix that maps the source keys $S_{[N]}$ to the individual keys $Z_{[K]}$. The individual key held by user $k$ is given by $Z_k = \mathbf{p}_k^T \mathbf{s}$. This scheme achieves the optimal communication rate $R_X^* = 1$ and the optimal total key rate $R_{Z_\Sigma}^* = N$ in \cite{yuan2025vector}.

In this framework, the distribution of individual key rates is entirely determined by the choice of the encoding matrix $\mathbf{P}$. The following lemma characterizes sufficient conditions on $\mathbf{P}$ under which the correctness and security constraints are satisfied.

\begin{lemma}\label{lemma-P}
    The scheme described in (\ref{eq:general_scheme}) satisfies correctness and security constraints if the encoding matrix $\mathbf{P}$ satisfies
    \begin{align}
        & \mathrm{Condition \ 1:} \ \mathbf{F} \mathbf{P} = \mathbf{0}_{M \times N}, \label{eq:cond1} \\
        & \mathrm{Condition \ 2:} \ \mathrm{rank}(\mathbf{G} \mathbf{P})  = N. \label{eq:cond2}
    \end{align}
\end{lemma}

\begin{Proof}
    For correctness, the server processes the messages $\mathbf{x}$ as $\mathbf{F} \mathbf{x} = \mathbf{F} \mathbf{w} + \mathbf{F}\mathbf{P} \mathbf{x} = \mathbf{F} \mathbf{w}$, which follows from $\mathrm{Condition}\ 1$ in (\ref{eq:cond1}).
    For the security constraint, we have
    \begin{align}
        & I(\mathbf{G}\mathbf{w}; X_{[K]} | \mathbf{F} \mathbf{w}) \nonumber\\
        &= H(X_{[K]} | \mathbf{F} \mathbf{w}) - H(X_{[K]} | \mathbf{F} \mathbf{w}, \mathbf{G}\mathbf{w}) \\
        & = H(X_{[K]}, \mathbf{F} \mathbf{w}) - H(\mathbf{F}\mathbf{w}) - H(X_{[K]}, \mathbf{G}\mathbf{x} | \mathbf{F} \mathbf{w}, \mathbf{G}\mathbf{w}) \\
        & = K - M - H(X_{[K]}, \mathbf{G}\mathbf{w} + \mathbf{G}\mathbf{P} \mathbf{s} | \mathbf{F} \mathbf{w}, \mathbf{G}\mathbf{w}) \label{ineq:1} \\
        & = K - M - H(X_{[K]}, \mathbf{G}\mathbf{P}\mathbf{s} | \mathbf{F} \mathbf{w}, \mathbf{G}\mathbf{w}) \\
        & = K - M - H(\mathbf{s}) - H(W_{[K]}| \mathbf{F} \mathbf{w}, \mathbf{G}\mathbf{w}) \label{eq:z} \\
        & = K - M - N - (K - M - N) = 0,
    \end{align}
    where the equality in (\ref{ineq:1}) holds because $H(X_{[K]},\mathbf{F}\mathbf{w}) = H(X_{[K]}) + H(\mathbf{F}\mathbf{w} | X_{[K]}) = K$, and (\ref{eq:z}) is because of the $\mathrm{Condition} \ 2$ in (\ref{eq:cond2}).
\end{Proof}

\begin{remark}
    Recall that the individual key held by user $k$ is given by $Z_k = \mathbf{p}_k^T \mathbf{s}$, where $\mathbf{p}_k^T$ is the $k$th row of $\mathbf{P}$. Hence, the individual key rate for user $k$ is determined by 
    \begin{align}
        R_k = H(Z_k)/L = \mathrm{rank}(\mathbf{p}_k^T) = \mathbbm{1}(\mathbf{p}_k \neq \mathbf{0}).
    \end{align}
    Thus, the presence of all-zero rows in $\mathbf{P}$ directly affects the individual key rate tuple. Therefore, the design of $\mathbf{P}$ can be equivalently viewed as determining which users hold non-zero individual keys, under the constraints that the correctness and security conditions are satisfied.
    In the following subsection, we establish a precise connection between feasible $\mathbf{P}$ and the rank-increment condition introduced earlier.
\end{remark}

\subsection{Design of the Encoding Matrix}\label{Sec:encoder}
We now show that for any index set $\mathcal{I} \subseteq [K] $ satisfying the rank-increment condition, there exists an encoding matrix $\mathbf{P}$ whose rows indexed by $\mathcal{I}$ are non-zero, while all remaining rows are zeros.

\begin{lemma}\label{lemma-encoder}
    For any index set $\mathcal{I} \subseteq [K]$ satisfying $\mathrm{rank}([\mathbf{F}_\mathcal{I};\mathbf{G}_\mathcal{I}]) = \mathrm{rank}(\mathbf{F}_\mathcal{I}) + N$, there exists a matrix $\bar{\mathbf{P}} \in \mathbb{F}_q^{|\mathcal{I}| \times N}$ such that
    \begin{align}
        \mathbf{F}_{\mathcal{I}} \bar{\mathbf{P}} = \mathbf{0}_{M \times N}, \quad \mathrm{rank}(\mathbf{G}_{\mathcal{I}} \bar{\mathbf{P}}) = N.
    \end{align}
\end{lemma}

The proof of Lemma~\ref{lemma-encoder} is given in Appendix~\ref{proof:lemma-encoder}.

\begin{remark}
    This lemma shows that whenever $\mathcal{I}$ satisfies the rank-increment condition, one can construct an encoding matrix $\mathbf{P}$ whose rows indexed by $\mathcal{I}$ are non-zero and whose remaining rows are identically zero, while still satisfying the correctness and security conditions.
    Such $\mathbf{P}$ corresponds to the individual key rate tuple $ (R_1,\cdots,R_K) =  (\mathbbm{1}(1\in \mathcal{I}), \cdots,\mathbbm{1}(K\in \mathcal{I}))$. Therefore, each such tuple is achievable, and the entire polyhedral region obtained by time-sharing among these tuples is achievable as well.
\end{remark}

\appendix 

\subsection{Proof of Theorem~\ref{thm-I*}} \label{proof:thm-I*}
We now prove the equality in (\ref{eq:size_equal}) by contradiction. Suppose that an inclusion-wise minimal subset $\mathcal{I}^*$ satisfies $\mathrm{rank}([\mathbf{F}_{\mathcal{I}^*};\mathbf{G}_{\mathcal{I}^*}]) = N + \mathrm{rank}(\mathbf{F}_{\mathcal{I}^*}) $ but $|\mathcal{I}^*| > N + \mathrm{rank}(\mathbf{F}_{\mathcal{I}^*})$. Then, the columns of $[\mathbf{F}_{\mathcal{I}^*};\mathbf{G}_{\mathcal{I}^*}]$ are linearly dependent. Consequently, there exists a nonempty subset $\mathcal{J} \subseteq \mathcal{I}^*$ such that
\begin{align}
    \mathrm{rank}([\mathbf{F}_{\mathcal{I}^*\setminus \mathcal{J}};\mathbf{G}_{\mathcal{I}^*\setminus \mathcal{J}}]) = N + \mathrm{rank}(\mathbf{F}_{\mathcal{I}^*}).
\end{align}
Since
\begin{align}
    \mathrm{rank}([\mathbf{F}_{\mathcal{I}^*\setminus \mathcal{J}};\mathbf{G}_{\mathcal{I}^*\setminus \mathcal{J}}]) & \leq N + \mathrm{rank}(\mathbf{F}_{\mathcal{I}^*\setminus \mathcal{J}})\\
    & \leq  N + \mathrm{rank}(\mathbf{F}_{\mathcal{I}^*}),
\end{align}
it follows that 
\begin{align}
    \mathrm{rank}([\mathbf{F}_{\mathcal{I}^*\setminus \mathcal{J}};\mathbf{G}_{\mathcal{I}^*\setminus \mathcal{J}}]) = N + \mathrm{rank}(\mathbf{F}_{\mathcal{I}^*\setminus \mathcal{J}}).
\end{align}
This implies that the set $\mathcal{I}^*\setminus \mathcal{J}$ also satisfies the rank-increment condition. This contradicts the assumption that $\mathcal{I}^*$ is inclusion-wise minimal. Hence, equality in (\ref{eq:size_equal}) must hold.

\subsection{Proof of Theorem~\ref{thm:2}} \label{proof:thm-2}
We begin by proving the first part of the theorem.
For any data stream length $L$, define $\mathbf{W} \defeq [W_1;\cdots;W_K] \in \mathbb{F}_q^{K \times L}$.
When only users in the index set $\mathcal{I}$ hold keys, for every $\ell \in [K]\setminus \mathcal{I}$, the transmitted symbol $X_{\ell}$ is a deterministic function of $W_\ell$, and hence
\begin{align}
    H(X_\ell |W_{\ell}) = 0, \quad \forall \ell \in [K]\setminus \mathcal{I}.
\end{align}
Moreover, it is shown in \cite{yuan2025vector} that $H(X_{\ell}) \geq L$.  Since $H(W_\ell) = L$, we have have
\begin{align}
    L \leq H(X_{\ell}) - H(X_\ell |W_{\ell}) =  H(W_{\ell}) - H(W_\ell |X_{\ell}) \leq L,
\end{align}
which implies $H(W_\ell | X_\ell) = 0, \ \forall \ell \in [K]\setminus \mathcal{I}$.
Now, by the security constraint,
\begin{align}
    I(\mathbf{G}\mathbf{W};X_{[K]} | \mathbf{F}\mathbf{W}) = NL - H(\mathbf{G}\mathbf{W}|X_{[K]}, \mathbf{F}\mathbf{W}) = 0.
\end{align}
Thus,
\begin{align}
     NL & = H(\mathbf{G}\mathbf{W}|X_{[K]}, \mathbf{F}\mathbf{W}) \label{eq:X_00}\\
     & = H(\mathbf{G}\mathbf{W}|X_{\mathcal{I}},W_{[K]\setminus \mathcal{I}}, \mathbf{F}_\mathcal{I}\mathbf{W}_{\mathcal{I},:}) \label{eq:X_W0} \\
     & \leq H(\mathbf{G}_{\mathcal{I}} \mathbf{W}_{\mathcal{I},:}| \mathbf{F}_{\mathcal{I}}\mathbf{W}_{\mathcal{I},:} ) \leq NL, \label{eq:X_01}
\end{align}
where $\mathbf{W}_{\mathcal{I},:}$ denotes the submatrix formed by the rows indexed by $\mathcal{I}$. It follows that
\begin{align}
    H(\mathbf{G}_\mathcal{I} \mathbf{W}_{\mathcal{I},:} |  \mathbf{F}_{\mathcal{I}} \mathbf{W}_{\mathcal{I},:} ) = NL. \label{eq:G_X}
\end{align}
Since $\mathbf{W}_{\mathcal{I},:}$ is uniformly distributed over $\mathbb{F}_q$, the above equality is equivalent to
\begin{align}
    \mathrm{rank}([\mathbf{F}_{\mathcal{I}};\mathbf{G}_{\mathcal{I}}]) = N + \mathrm{rank}(\mathbf{F}_{\mathcal{I}}).
\end{align}
i.e., the rank-increment condition holds for $\mathcal{I}$.

We now prove the second part of the theorem. Suppose that $\mathcal{I}^* \subseteq [K]$ satisfies the rank-increment condition and is inclusion-wise minimal.
For any $i \in \mathcal{I}^*$, define $\mathcal{I}' \defeq \mathcal{I}^* \setminus \{i\}$. By the minimality of $\mathcal{I}^*$, 
\begin{align}
    \mathrm{rank}([\mathbf{F}_{\mathcal{I}'}; \mathbf{G}_{\mathcal{I}'}]) < \mathrm{rank}(\mathbf{F}_{\mathcal{I}'}) + N.
\end{align}
Recalling the equivalent characterization of rank-increment condition in (\ref{eq:equiv_cond}), at least one of the following two cases occurs:
\begin{itemize}
    \item \textbf{Case 1}: $\mathrm{rowspan}(\mathbf{F}_{\mathcal{I}'}) \cap \mathrm{rowspan}(\mathbf{G}_{\mathcal{I}'}) \neq \{0\}$.

    \item \textbf{Case 2}: $\mathrm{rank}(\mathbf{G}_{\mathcal{I}'}) \leq N-1$.
\end{itemize}
We next show that in either case, we have $R_i \geq 1$.

In \textbf{Case 1}, there exist nonzero vectors $\mathbf{b} \in \mathbb{F}_q^{N \times 1}$ and $\mathbf{c}  \in \mathbb{F}_q^{M \times 1} $ such that
\begin{align}
    \mathbf{b}^T \mathbf{G}_{\mathcal{I}'} =  \mathbf{c}^T \mathbf{F}_{\mathcal{I}'} \defeq \mathbf{a}^T \neq \mathbf{0}, \quad  \mathbf{b}^T \mathbf{G}_{\mathcal{I}^*} \neq \mathbf{c}^T \mathbf{F}_{\mathcal{I}^*},
\end{align}
which also implies
\begin{align}
    \mathbf{b}^T \mathbf{G}_{\{i\}} \neq \mathbf{c}^T \mathbf{F}_{\{i\}}. \label{eq:b_c}
\end{align}
Then, from (\ref{eq:X_00})--(\ref{eq:X_01}), we have
\begin{align}
    NL & = H(\mathbf{G}\mathbf{W} | \mathbf{F}_{\mathcal{I}^*} \mathbf{W}_{\mathcal{I}^*,:},X_{\mathcal{I}^*}, W_{[K] \setminus \mathcal{I}^*}) \\
    & = H(\mathbf{G}_{\mathcal{I}^*} \mathbf{W}_{\mathcal{I}^*,:} | \mathbf{F}_{\mathcal{I}^*} \mathbf{W}_{\mathcal{I}^*,:},X_{\mathcal{I}^*}, W_{[K] \setminus \mathcal{I}^*}) \\
    & \leq H(\mathbf{G}_{\mathcal{I}^*} \mathbf{W}_{\mathcal{I}^*,:} | \mathbf{F}_{\mathcal{I}^*} \mathbf{W}_{\mathcal{I}^*,:},X_{\mathcal{I}^*}) \leq NL,
\end{align}
which implies
\begin{align}
    H(\mathbf{G}_{\mathcal{I}^*} \mathbf{W}_{\mathcal{I}^*,:} | \mathbf{F}_{\mathcal{I}^*} \mathbf{W}_{\mathcal{I}^*,:},X_{\mathcal{I}^*}) = NL.
\end{align}
Hence,
\begin{align}
    & I(\mathbf{G}_{\mathcal{I}^*} \mathbf{W}_{\mathcal{I}^*,:}; \mathbf{F}_{\mathcal{I}^*} \mathbf{W}_{\mathcal{I}^*,:},X_{\mathcal{I}^*}) \nonumber \\
    & = H(\mathbf{G}_{\mathcal{I}^*} \mathbf{W}_{\mathcal{I}^*,:}) - H(\mathbf{G}_{\mathcal{I}^*} \mathbf{W}_{\mathcal{I}^*,:} | \mathbf{F}_{\mathcal{I}^*} \mathbf{W}_{\mathcal{I}^*,:},X_{\mathcal{I}^*}) \\
    & = H(\mathbf{G}_{\mathcal{I}^*} \mathbf{W}_{\mathcal{I}^*,:}) - NL = 0.
\end{align}
Since $i \in \mathcal{I}^*$, we have $I(\mathbf{G}_{\mathcal{I}^*} \mathbf{W}_{\mathcal{I}^*,:}; \mathbf{F}_{\mathcal{I}^*} \mathbf{W}_{\mathcal{I}^*,:},X_{i}) = 0$, and therefore,
\begin{align}
    0 & = I(\mathbf{b}^T \mathbf{G}_{\mathcal{I}^*} \mathbf{W}_{\mathcal{I}^*,:}; \mathbf{c}^T \mathbf{F}_{\mathcal{I}^*} \mathbf{W}_{\mathcal{I}^*,:},X_{i}) \\
    & = I(\mathbf{a}^T \mathbf{W}_{\mathcal{I}',:} + \mathbf{b}^T \mathbf{G}_{\{i\}} W_{i}; \mathbf{a}^T \mathbf{W}_{\mathcal{I}',:} + \mathbf{c}^T \mathbf{F}_{\{i\}} W_{i},X_{i}).
\end{align}
Consequently, we have
\begin{align}
    I(\mathbf{a}^T \mathbf{W}_{\mathcal{I}',:} + \mathbf{b}^T \mathbf{G}_{\{i\}} W_{i}; X_{i} | \mathbf{a}^T \mathbf{W}_{\mathcal{I}',:} + \mathbf{c}^T \mathbf{F}_{\{i\}} W_{i}) = 0.
\end{align}
It follows that
\begin{align}
    & H((\mathbf{b}^T \mathbf{G}_{\{i\}} - \mathbf{c}^T \mathbf{F}_{\{i\}}) W_i | \mathbf{a}^T \mathbf{W}_{\mathcal{I}',:} + \mathbf{c}^T \mathbf{F}_{\{i\}} W_i) \notag \\
    & - H((\mathbf{b}^T \mathbf{G}_{\{i\}} - \mathbf{c}^T \mathbf{F}_{\{i\}}) W_i |X_{i}, \mathbf{a}^T \mathbf{W}_{\mathcal{I}',:} + \mathbf{c}^T \mathbf{F}_{\{i\}} W_i) \nonumber \\
    & = L - H((\mathbf{b}^T \mathbf{G}_{\{i\}} - \mathbf{c}^T \mathbf{F}_{\{i\}}) W_i |X_{i}, \mathbf{a}^T \mathbf{W}_{\mathcal{I}',:} + \mathbf{c}^T \mathbf{F}_{\{i\}} W_i) \\
    & = 0,
\end{align}
which implies
\begin{align}
    L & = H((\mathbf{b}^T \mathbf{G}_{\{i\}} - \mathbf{c}^T \mathbf{F}_{\{i\}}) W_i |X_{i}, \mathbf{a}^T \mathbf{W}_{\mathcal{I}',:} + \mathbf{c}^T \mathbf{F}_{\{i\}} W_i) \\
    & \leq H((\mathbf{b}^T \mathbf{G}_{\{i\}} - \mathbf{c}^T \mathbf{F}_{\{i\}}) W_{i} |X_{i}) \leq L.
\end{align}
Recall that in (\ref{eq:b_c}), $\mathbf{b}^T \mathbf{G}_{\{i\}} - \mathbf{c}^T \mathbf{F}_{\{i\}} \neq 0$, and then $H(W_i|X_i) = L$.
Thus,
\begin{align}
    H(Z_i) & \geq H(X_i | W_i) \label{eq:Z}\\
    & = H(X_i) + H(W_i | X_i) - H(W_i) \\
    & = H(X_i) \geq L,
\end{align}
where (\ref{eq:Z}) is because $H(X_i | W_i, Z_i) = 0$.
Hence, $R_i = L_i/L \geq H(Z_i)/L \geq 1$.

In \textbf{Case 2}, from (\ref{eq:G_X}), we have
\begin{align}
    NL & = H(\mathbf{G}_{\mathcal{I}^*} \mathbf{W}_{\mathcal{I}^*,:} | X_{\mathcal{I}^*}) \\
    &\leq H(\mathbf{G}_{\mathcal{I}'} \mathbf{W}_{\mathcal{I}',:} |X_{\mathcal{I}^*}) + H(\mathbf{G}_{\{i\}} W_{i }|X_{\mathcal{I}^*}) \\
    & \leq (N-1)L + H(\mathbf{G}_{\{i\}} W_{i} |X_{i}) \\
    & \leq (N-1)L + H(W_i |X_{i}),
\end{align}
and then
\begin{align}
    R_i \geq H(W_i |X_{i})/L \geq 1.
\end{align}

Thus, both cases lead to $R_i \geq 1, \forall i \in \mathcal{I}^*$, completing the proof.

\subsection{Proof of Lemma~\ref{lemma-encoder}}\label{proof:lemma-encoder}
Let $r \defeq \mathrm{rank}(\mathbf{F}_\mathcal{I})$. The nullspace of $\mathbf{F}_\mathcal{I}$ is defined as
\begin{align}
    \mathcal{N}(\mathbf{F}_\mathcal{I}) \defeq \{ \mathbf{x} \in \mathbb{F}_q^{|\mathcal{I}|} | \mathbf{F}_\mathcal{I} \mathbf{x} = \mathbf{0}\}.
\end{align}
Let $\mathbf{U} \in \mathbb{F}_q^{|\mathcal{I}| \times (|\mathcal{I}|-r)} $ be any matrix whose columns form a basis of $\mathcal{N}(\mathbf{F}_\mathcal{I})$. This implies $\mathbf{F}_\mathcal{I} \mathbf{U} = \mathbf{0}$.
Hence, for any matrix $\bar{\mathbf{P}}$ whose columns lie in $\mathcal{N}(\mathbf{F}_{\mathcal{I}})$, we have $\mathbf{F}_{\mathcal{I}} \bar{\mathbf{P}} = \mathbf{0}$. It remains to show that we can choose $N$ such columns such that $\mathrm{rank}(\mathbf{G}_{\mathcal{I}} \bar{\mathbf{P}}) = N$.
We first establish the following rank decomposition identity.
\begin{claim}\label{claim-1}
    For any matrices $\mathbf{A} \in \mathbb{F}_q^{a \times m}$ and $\mathbf{B} \in \mathbb{F}_q^{b \times m}$, 
    \begin{align}
        \mathrm{rank}([\mathbf{A};\mathbf{B}]) = \mathrm{rank}(\mathbf{A}) + \mathrm{dim}(\mathbf{B} \mathcal{N}(\mathbf{A})),
    \end{align}
    where $\mathbf{B} \mathcal{N}(\mathbf{A}) \defeq \{ \mathbf{B}\mathbf{x}|\mathbf{x} \in \mathcal{N}(\mathbf{A}) \}$.
\end{claim}
\begin{Proof}
    Consider the linear map $T: \mathbb{F}_q^m \to \mathbb{F}_q^{a+b}$ defined by
    \begin{align}
        T(\mathbf{x}) = [\mathbf{A};\mathbf{B}]\mathbf{x} = \begin{bmatrix} \mathbf{A}\mathbf{x} \\ \mathbf{B}\mathbf{x} \end{bmatrix}.
    \end{align}
    Let $V$ denote the image of $T$, i.e.,
    \begin{align}
        V \defeq \mathrm{Im}(T) = \left\{ \begin{bmatrix} \mathbf{A}\mathbf{x} \\ \mathbf{B}\mathbf{x} \end{bmatrix} \;\middle|\; \mathbf{x} \in \mathbb{F}_q^m \right\}.
    \end{align}
    Define the projection map $\pi: V \to \mathrm{Im}(\mathbf{A})$ by
    \begin{align}
        \pi\left( \begin{bmatrix} \mathbf{A}\mathbf{x} \\ \mathbf{B}\mathbf{x} \end{bmatrix} \right) = \mathbf{A}\mathbf{x}.
    \end{align}
    Thus, $\pi$ is linear and surjective, and its kernel is
    \begin{align}
        \ker(\pi) = \left\{ \begin{pmatrix} \mathbf{A}\mathbf{x} \\ \mathbf{B}\mathbf{x} \end{pmatrix} \in V \;\middle|\; \mathbf{A}\mathbf{x} = \mathbf{0} \right\}
    = \left\{ \begin{bmatrix} \mathbf{0} \\ \mathbf{B}\mathbf{x} \end{bmatrix} \;\middle|\; \mathbf{x} \in \mathcal{N}(\mathbf{A}) \right\}.
    \end{align}
    Now, define a linear map $\phi: \mathbf{B}\mathcal{N}(\mathbf{A}) \to \ker(\pi)$ by 
    \begin{align}
        \phi(\mathbf{B}\mathbf{x}) = \begin{bmatrix} \mathbf{0} \\ \mathbf{B}\mathbf{x} \end{bmatrix}, \quad \mathbf{x} \in \mathcal{N}(\mathbf{A}).
    \end{align}
    The map $\phi$ is well-defined, linear, injective, and surjective. Hence, $\phi$ is an isomorphism, and
    \begin{align}
        \mathrm{dim}(\ker(\pi)) = \mathrm{dim}(\mathbf{B}\mathcal{N}(\mathbf{A})).
    \end{align}
    Applying the rank-nullity theorem~\cite{strang2012linear}, we obtain
    \begin{align}
        \mathrm{dim}(V) & = \mathrm{dim}(\mathrm{Im}(\pi)) + \mathrm{dim}(\ker(\pi)) \\
        & = \mathrm{dim}(\mathrm{Im}(\mathbf{A})) + \mathrm{dim}(\mathbf{B}\mathcal{N}(\mathbf{A})).
    \end{align}
    It follows that
    \begin{align}
        \mathrm{rank}([\mathbf{A};\mathbf{B}]) = \mathrm{rank}(\mathbf{A}) + \mathrm{dim}\left(\mathbf{B} \mathcal{N}(\mathbf{A})\right),
    \end{align}
    which proves the claim.
\end{Proof}

Now, applying the claim~\ref{claim-1}, we have
\begin{align}
    \mathrm{rank}([\mathbf{F}_{\mathcal{I}};\mathbf{G}_{\mathcal{I}}]) = \mathrm{rank}(\mathbf{F}_{\mathcal{I}}) + \mathrm{dim}(\mathbf{G}_{\mathcal{I}} \mathcal{N}(\mathbf{F}_{\mathcal{I}})).
\end{align}
This leads to $\mathrm{dim}(\mathbf{G}_{\mathcal{I}} \mathcal{N}(\mathbf{F}_{\mathcal{I}})) = N$ by the assumption in Lemma~\ref{lemma-encoder}. Since $\mathrm{colspan}(\mathbf{U}) = \mathcal{N}(\mathbf{F}_{\mathcal{I}})$, we have $\mathbf{G}_{\mathcal{I}} \mathcal{N}(\mathbf{F}_{\mathcal{I}}) = \mathrm{colspan}(\mathbf{G}_{\mathcal{I}} \mathbf{U})$, and hence, $\mathrm{rank}(\mathbf{G}_{\mathcal{I}} \mathbf{U}) = N$.
Therefore, $\mathbf{G}_{\mathcal{I}} \mathbf{U}$ has $N$ independent columns, indexed by $j_1,\ldots,j_N$. We define
\begin{align}
    \bar{\mathbf{P}} \defeq [\mathbf{U}_{:,j_1} \cdots \mathbf{U}_{:,j_N}] \in \mathbb{F}_q^{|\mathcal{I}| \times N}.
\end{align}
Then, $\mathrm{rank}(\mathbf{G}_{\mathcal{I}} \bar{\mathbf{P}})=N$ and $\mathbf{F}_\mathcal{I} \bar{\mathbf{P}} = \mathbf{0}$ since $\bar{\mathbf{P}}$ consists of columns from $\mathcal{N}(\mathbf{F}_{\mathcal{I}})$, which proves Lemma~\ref{lemma-encoder}.

\bibliographystyle{unsrt}
\bibliography{Ref1.bib}

\end{document}